\documentclass[aps,pre,preprint,superscriptaddress,showpacs,showkeys,11pt,a4paper]{revtex4}
\usepackage{amsfonts}
\usepackage{amssymb}
\usepackage{amsmath}
\usepackage{graphicx}
\usepackage{color}

\begin{document} 

\title{Mechano-sysntesis and structural, magnetic and thermal characterization of $\alpha$-Fe nanoparticles embedded in a W\"{u}stite matrix}

\author{S. O. S. Batista}
\affiliation{Departamento de F\'{i}sica Te\'{o}rica e Experimental, Universidade Federal do Rio Grande do Norte, 59078-900 Natal, RN, Brazil} 
\author{M. A. Morales}
\email[Electronic address: ]{morales@dfte.ufrn.br}
\affiliation{Departamento de F\'{i}sica Te\'{o}rica e Experimental, Universidade Federal do Rio Grande do Norte, 59078-900 Natal, RN, Brazil} 
\author{W. C. dos Santos}
\affiliation{Departamento de F\'{i}sica Te\'{o}rica e Experimental, Universidade Federal do Rio Grande do Norte, 59078-900 Natal, RN, Brazil} 
\author{C. A. Iglesias}
\affiliation{Departamento de F\'{i}sica Te\'{o}rica e Experimental, Universidade Federal do Rio Grande do Norte, 59078-900 Natal, RN, Brazil} 
\author{E. Baggio-Saitovitch}
\affiliation{Centro Brasileiro de Pesquisas F\'{i}sicas, Rua Dr.\ Xavier Sigaud 150, Urca, 22290-180 Rio de Janeiro, RJ, Brazil} 
\author{A. S. Carri\c{c}o}
\affiliation{Departamento de F\'{i}sica Te\'{o}rica e Experimental, Universidade Federal do Rio Grande do Norte, 59078-900 Natal, RN, Brazil} 
\author{F. Bohn} 
\email[Electronic address: ]{felipebohn@gmail.com}
\affiliation{Escola de Ci\^{e}ncias e Tecnologia, Universidade Federal do Rio Grande do Norte, 59078-900 Natal, RN, Brazil} 
\affiliation{Departamento de F\'{i}sica Te\'{o}rica e Experimental, Universidade Federal do Rio Grande do Norte, 59078-900 Natal, RN, Brazil} 
\author{S. N. de Medeiros}
\email[Electronic address: ]{snm@dfte.ufrn.br}
\affiliation{Departamento de F\'{i}sica Te\'{o}rica e Experimental, Universidade Federal do Rio Grande do Norte, 59078-900 Natal, RN, Brazil} 

\date{\today} 

\begin{abstract} 
Magnetic materials for specific applications require an accurate control and complete comprehension of their magnetic properties. In particular, nanoparticles embedded in a polycrystalline matrix emerge as good candidates for applications due to the possibility of tuning the magnetic properties through interface interaction effects. Here, iron/w\"{u}stite composite is prepared using high energy mechanical milling from iron powder and water. The sample is analyzed by X-ray diffraction, dynamic laser light scattering, M\"{o}ssbauer spectroscopy, field cooling and zero field cooling curves, magnetization curves, and magnetic hyperthermia. Based on the results, we identify that the produced sample is like Fe nanoparticles embedded in a w\"{u}stite matrix, with high stability in time, and shows noticeable features such as exchange bias effect at low temperatures and promising temperatures reached in a short time interval when considered magnetic hyperthermia, $\sim 46\,^\circ$C, becoming an interesting candidate for biological applications, such as the one employed for cancer therapy.
\end{abstract} 

\pacs{75.75.-c, 75.75.Jn, 65.80.-g, 82.60.Qr}

\keywords{Iron, W\"{u}stite, M\"{o}ssbauer, Hyperthermia} 

\maketitle 

\section{Introduction} 
\label{Introduction}

Mechanical alloying/milling is a non-equilibrium process and corresponds to a widely used low cost technique to produce metaestable alloy phases, amorphous phases and intermetallics compounds from crystalline elementary powder mixtures. As a remarkable point, in the case when a single elementary powder is employed, mechanical milling is able to produce nanoparticles~\cite{PMS42p311, JMS39p5045}, although it involves a complex process and requires optimization of a few parameters to obtain the required phase, absence of residues and control of the particle size. 

In general, nanoparticles (NPs) are of great interest because they can be employed in a wide range of technological applications, from magnetic data storage to chemical and biomedical uses. Considering biological applications, the primary interest resides in the therapeutic potential of heat they can generate, since magnetic hyperthermia is a promising approach to cancer therapy. In particular, the well-known therapeutic procedure named hyperthermia corresponds simply to raising the temperature of a tumor up to $42 - 45\,^\circ$C, causing its necrosis without overheating and damaging the surrounding normal tissue. With this spirit, magnetic hyperthermia becomes an important issue because the ferromagnetic NPs can generate heat at a localized area by different mechanisms (hysteresis, eddy currents, N\'{e}el switching, and Brownian rotation) when submitted to an external alternating magnetic field~\cite{N7p1443, RPOR18p397}. For this reason, the complete characterization of the particles becomes fundamental to obtain further information on the nature of the heating and dynamical magnetic behavior of the nanoparticles.

In this respect, several materials have already been studied and tested. In particular, most of them have low saturation magnetization, like ferri/ferromagnetic iron oxides NPs~\cite{N7p1443}. However, the increase of magnetization provided by the metallic Fe NPs allows possibly more effective generation of local heat than, e.\ g., ferrimagnetic magnetite or maghemite, when subjected to an alternating magnetic field in a frequency range from $50$ kHz to a few hundred kHz. 

In this sense, Fe nanoparticles are of great interest in technological applications due to properties such as high saturation magnetization value at room temperature and high reactivity. In fact, the last one can be a drawback for its application since Fe NPs are immediately oxidized through an exothermic reaction with air or water traces. Therefore, their synthesis must be conducted under controlled atmosphere. To reduce or avoid problems with respect to the high reactivity of the Fe particles, passivation of the surface by an iron oxide layer has appeared as a valuable approach for the stabilization of such NPs. The FeO phase is thermodynamically stable only above $843$ K and under low partial pressures of oxygen. Below $843$ K, the iron oxide decomposes into iron and magnetite. At low temperatures, there is the propensity that the iron is trivalent, avoiding the formation of stoichiometric FeO when quenched to temperatures of $\sim 300$ K, in which it is metaestable with long life~\cite{JAP37p3043}.

In this paper, we report an experimental study on Fe NPs passivated in a w\"{u}stite matrix prepared using high energy mechanical milling, in which the properties are investigated by X-ray diffraction, dynamic laser light scattering, M\"{o}ssbauer spectroscopy, field cooling and zero field cooling curves, magnetization curves, and magnetic hyperthermia. We identify that the produced sample is like Fe nanoparticles embedded in a w\"{u}stite matrix, with high stability in time, and shows noticeable features such as exchange bias effect at low temperatures and promising temperatures reached in a short time interval when considered magnetic hyperthermia, becoming an interesting candidate for biological applications, such as the one employed for cancer therapy.

\section{Experiment}
\label{Experiment}

\subsection{Sample preparation}
\label{Sample_preparation}

The studied sample, named FeM40h, is a powder prepared using a planetary high energy mechanical milling Fritsch Pulverisette 7 - Premium line system. The precursor material employed to produce the sample is Fe powder ($99.99\%$ purity) from Sigma - Aldrich. The masses considered for the synthesis are $0.8288$ g and $0.3562$ g for Fe and water, respectively. The milling is performed with a hardened steel vial with a total volume of $50$ mL. The ball to powder mass ratio is of $20:1$ and the rotation speed is of $600$ rpm. The balls and powder are deposited in the vial in normal atmosphere and are effectively milled during a time period of $40$ h, with standby time of $10$ min between each hour. The vial remains hermetically closed until reaching the whole milling time. At the end of the 40th hour milling time, the cap's vial is removed and it is allowed the sample to cool down to room temperature. 

\subsection{Characterization}
\label{Characterization}

Phase characterization of the produced powder is carried out via X-ray diffraction (XRD) using a Rigaku MiniFlex II diffractometer, with Cu K$\alpha$ radiation of $\lambda = 0.154$ nm. Diffraction measurement is performed in the angular range of $20^\circ \leq 2\theta \leq 90^\circ$, and the obtained XRD pattern is refined by Rietveld method.

For measuring the particle size of the NPs, dynamic laser light scattering is employed. The sample is dispersed in water, $0.1$ mg/mL, using a water-bath sonicator for $2$ min and the particle size is measured using a glass cuvette Brookhaven Instruments Corporation ZetaPlus potential analyzer. 

The $^{57}$Fe M\"{o}ssbauer characterization is performed in the transmission mode in a conventional constant acceleration spectrometer from Wiessel and using a $^{57}$Co(Rh) gamma ray source with activity of $25$ mCi. The M\"{o}ssbauer spectroscopy (MS) measurements are performed at room temperature. Isomer shifts (ISs) are reported having as reference the IS from $\alpha$-Fe at room temperature.

Magnetic characterization is obtained through zero field cooling (ZFC) and field cooling (FC) magnetization measurements as a function of the temperature, from $50$ K to $300$ K, as well as FC magnetization ($M\times H$) curves acquired at selected temperatures, with a Quantum Design model VersaLabTM vibrating sample magnetometer. In the FC magnetization measurements as a function of the temperature, a DC magnetic field of $100$ Oe is applied, while for the FC $M\times H$ magnetization curves, the sample is submitted to a DC magnetic field of $10$ kOe and cooled down from $300$ K to the set temperature.

Finally, thermal characterization of the produced sample is performed via magnetic hyperthermia (MH) using a home-made system. The sample is submitted to an alternating magnetic field with frequency of $58$ kHz and selected maximum amplitudes in the range between $90$ and $210$ Oe. The temperature data acquisition is performed using an infrared thermometer with sampling rate of $1$ sample per second. The measurements are carried out at room temperature. In this case, the sample is thermally isolated from the solenoid and the temperature changes due to resistive heating are negligible since the whole system responsible by the field generation is refrigerated.

\section{Results and discussion}
\label{Results_and_discussion}

In this section, we present the experimental results of the structural, magnetic and thermal characterization of the produced sample.

\subsection{Phase characterization}
\label{Phase_characterization}

\begin{figure}[!h] 
    \begin{center} 
    \includegraphics[width=7cm]{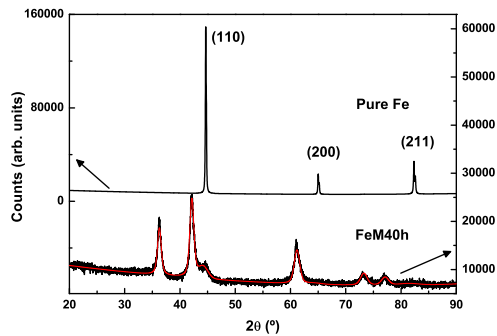} 
    \end{center} 
\vspace{-.5cm}
    \caption{XRD patterns for the samples of Fe powder and FeM40h sample, represented by the black lines. For the FeM40h sample, the red line is the experimental fit, obtained via Rietveld method.} 
    \label{Fig_01} 
\end{figure}

Figure \ref{Fig_01} shows the X-ray diffraction patterns obtained for the FeM40h sample and for the Fe powder precursor material employed in the synthesis. The diffractogram for the FeM40h sample is refined via the Rietveld method. From the fit, it is clear that the size and amount of the Fe powder decreases with the milling time, since after $40$ h of milling only the ($110$) Fe peak is present, having a very low intensity when compared to the iron oxide phase. The result indicates that the produced sample is a two phases system, composed by w\"{u}stite and $\alpha$-Fe, both with particles of reduced sizes. The FeM40h sample contains a fraction of $84$ w\% of $16$ nm w\"{u}stite particles and $16$ w\% of $8$ nm $\alpha$-Fe particles. In particular, the XRD pattern does not show any evidence of a third phase, namely magnetite or hematite. 

The lattice parameters $a$ for the $\alpha$-Fe and w\"{u}stite phases are of $0.2883$ nm and $0.4283$ nm, respectively. According to results previously published for single crystals of iron oxide, Fe$_y$O, with variable stoichiometry, the lattice parameter found in our sample is related to a $y = 0.893$, as calculated from the formula $a(y)=0.3856+0.0478y$, with $a$ in nm~\cite{PCM10p106}. The iron oxide has a NaCl-type crystal structure. Although the ideal w\"{u}stite uses to have cubic NaCl structure, stoichiometric FeO is unknown at normal pressure. The deviation from stoichiometry observed in our sample Fe$_{0.893}$O is related to cation vacancies and to ferric ions occupying octahedral and tetrahedral sites~\cite{IC23p3136}. The number of these defects is so massive that complex interactions between them lead to the formation of defect clusters~\cite{Cullity}.

As a remarkable point, after the first run of measurements and analysis, the sample was kept in capped plastic vials without any special enclosure. Several weeks later, the XRD measurements have been repeated and it is found that the newer pattern is identical to that previously obtained for the FeM40h sample, indicating that the relative mass concentrations for both phases are unchanged. This result shows that the w\"{u}stite phase is stable and may be protecting the $\alpha$-Fe phase from oxidation.

As previously mentioned, there are some missing Fe$^{2+}$ ions at the w\"{u}stite constitutional structure. To satisfy the charge neutrality, some ferrous ions oxidize to Fe$^{3+}$. This leads to the occupancy of Fe$^{3+}$ in both octahedral ($O$) and tetrahedral ($T$) interstitial sites~\cite{IC23p3136}. Thus, the chemical formula of w\"{u}stite can be written in terms of the iron occupancy as: 
\begin{equation}
\textrm{Fe}_{1-x}\textrm{O} = [\textrm{Fe}^{2+} _O]_{1-3x} [\textrm{Fe}^{3+} _O]_{2x-t} [\textrm{Fe}^{3+} _T]_t [\square _O] \textrm{O}^{2-}\,,
\end{equation}
\noindent in which $\square$ is an octahedral cation vacancy, while $x$ and $t$ measure the proportion of missing $O$ and $T$ iron ions.

\subsection{Dynamic laser light scattering}
\label{Dynamic_laser_light_scattering}

\begin{figure}[!h] 
    \begin{center} 
    \includegraphics[width=7cm]{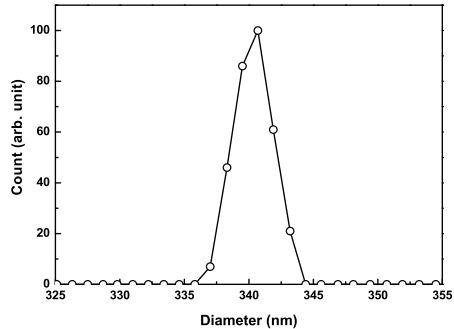} 
    \end{center} 
\vspace{-.5cm}
    \caption{Dynamic laser light scattering analysis for the FeM40h sample.} 
    \label{Fig_02} 
\end{figure}

Figure \ref{Fig_02} shows the dynamic laser light scattering analysis for the FeM40h sample. Whereas the $\alpha$-Fe and Fe$_{0.893}$O crystallite size calculated by analyzing the XRD peaks are of $8$ and $16$ nm, respectively, here, the obtained mean particle size is $340$ nm. In this case, the larger particle size measured through this technique, which indicates the hydrodynamic diameter, is associated to $\alpha$-Fe particles embedded in a polycrystalline matrix of Fe$_{0.893}$O. This may be a consequence of aggregation of the NPs, a fact commonly verified in milled samples due to successive processes of cold welding between the particles.

\subsection{M\"{o}ssbauer spectroscopy}
\label{Mossbauer_spectroscopy}

\begin{figure}[!h] 
    \begin{center} 
    \includegraphics[width=7cm]{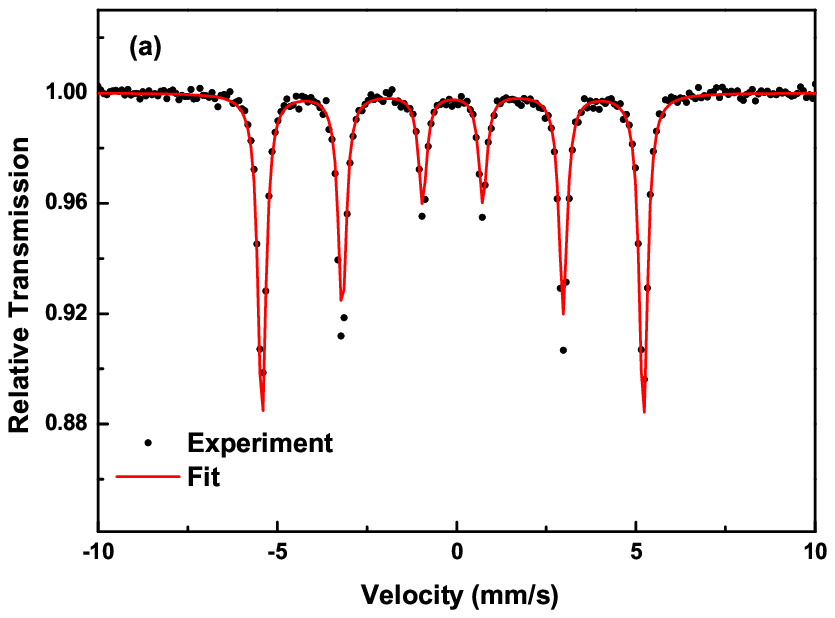}\\ 
    \includegraphics[width=7cm]{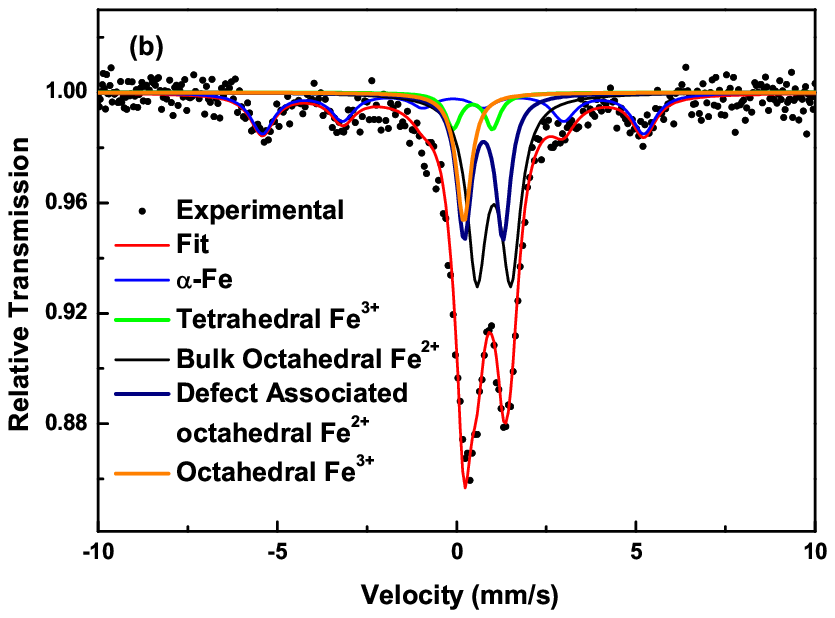}
    \end{center} 
\vspace{-.5cm}
    \caption{M\"{o}ssbauer spectrum measured  at room temperature for the (a) Fe powder and (b) FeM40h sample, together with the fit of the experimental data.} 
    \label{Fig_03} 
\end{figure}

M\"{o}ssbauer spectroscopy is an attractive technique to study the produced sample due to the presence of iron cations in different environments. Figure~\ref{Fig_03} shows the room temperature M\"{o}ssbauer spectrum obtained for the Fe powder and the FeM40h sample. For the $\alpha$-Fe powder, the MS shows only one component with hyperfine magnetic field (BHf) of $33$ T and IS of $0.0$ mm/s. In the case of the FeM40h sample, the MS reveals a spectrum that can be modeled with five subspectra: a sextet related to Fe, three doublets and a singlet related to w\"{u}stite. The sextet has similar hyperfine parameters as the pristine sample, a result in agreement with the one usually verified for magnetically blocked $\alpha$-Fe particles. On the other hand, w\"{u}stite is paramagnetic at room temperature and antiferromagnetically ordered at temperatures below $195$ K~\cite{Cullity}. 

In previous studies on Fe$_y$O samples, several physical models have been proposed to analyze the MS from samples with different $y$ values and prepared by sintering stoichiometric amounts of powdered Fe$_2$O$_3$ and $\alpha$-Fe. For samples with $0.95 < y < 1$~\cite{PCM11p250}, the fitting model consists in one singlet ascribed to Fe$^{3+}$ and two doublets related to Fe$^{2+}$ at different distorted environments in $O$ sites. For samples with $0.87 < y < 0.95$~\cite{IC23p3136, IC21p2804} the model consists in one singlet due to Fe$^{3+}$ in $O$ sites, two doublets associated to Fe$^{2+}$ in $O$ sites and one doublet attributed to Fe$^{3+}$ in $T$ site. Here, we adopt the fitting model reported in references~\cite{IC23p3136, IC21p2804}, where the singlet is associated to O ferric ions, the two doublets are related to O ferrous ions located near to vacancies and interstitials cations. The fourth subspectra, doublet, is associated to Fe$^{2+}$ ions occupying $T$ interstitial sites. 

The hyperfine parameters obtained from the fit are presented in Table~\ref{Table_01}. The values of Area-w\"{u}stite show the percentile of Fe distributed at different environments in the Fe$_{0.893}$O phase. These values are calculated from the absolute areas obtained from the fits. Substitutional octahedral Fe$^{3+}$ ions, which are provided to allow charge neutrality, are subjected to a crystal field with cubic symmetry and thus the electric field gradient (EFG) is zero at the iron nucleus~\cite{PCM11p250}. Since the Fe$^{2+}$ radi is larger than Fe$^{3+}$ ions, some of the Fe$^{3+}$ ions at $O$ sites may diffuse to interstitial $T$ sites. At these sites, the adjacent cation defects develop an EFG which splits the I$_{3/2}$ nuclear level, leading to a quadrupole interaction. The larger quadrupole splitting (QS) for Fe$^{3+}$ at $T$ site must result from a lattice contribution to the EFG. The Fe$^{2+}$ ions are located at $O$ sites and exhibit hyperfine parameters typical for high spin Fe$^{2+}$. The parameters obtained are similar to the results reported for Fe$^{2+}$ cations in a non cubic environment, probably due to vacancies as near neighborhoods of the Fe$^{2+}$ ions~\cite{IC21p2804}. The QS for the Fe$^{2+}$ cations are larger than the ones reported in~\cite{PCM11p250}, and this feature may be devoted to the high disorder degree induced by the milling process. The linewidths of all subespectra are broad, indicating that more than one site with slightly different parameter is represented by each component. The obtained relative absorption areas for the sextet and the subspectra ascribed to w\"{u}stite are respectively of $22$\% and $78$\%. This result indicates a reasonable agreement with the mass percentage obtained from the XRD analysis. 

\begin{table*}[!h]
\begin{center}
\caption{Hyperfine parameters for the FeM40h sample obtained from the fit of the M\"{o}ssbauer spectrum measured at room temperature. In this case, IS is the isomer shifts, QS is the quadrupole splitting, and BHf represents the hyperfine magnetic field.}
\label{Table_01}
\begin{tabular}{l c c c c c}
\hline \hline
 &\hspace{.1cm}IS (mm/s)\hspace{.1cm}&\hspace{.1cm} QS (mm/s) \hspace{.1cm}&\hspace{.1cm} Area (\%) \hspace{.1cm}&\hspace{.1cm} Area W\"{u}stite (\%) \hspace{.1cm}&\hspace{.1cm} BHf (T)\\ \hline
Bulk Octahedral, Fe$^{2+}$ & $1.14$ & $0.95$ & $38$ & $49$ & $-$\\ \hline
Defect associated octahedral, Fe$^{2+}$ & $0.86$ & $1.08$ & $23$ & $30$ & $-$\\ \hline
Octahedral, Fe$^{3+}$ & $0.31$ &$-$ &$11$ & $14$ &$-$\\ \hline
Tetrahedral, Fe$^{3+}$ & $0.56$ & $1.10$ & $6$ & $7$ & $-$ \\ \hline
$\alpha$-Fe & $0.0$ & $-$ & $22$ & & $33.0$\\
\hline \hline
\end{tabular}
\end{center}
\end{table*}

\subsection{Magnetic behavior}
\label{Magnetic behavior}

\begin{figure}[!h] 
    \begin{center} 
    \includegraphics[width=7cm]{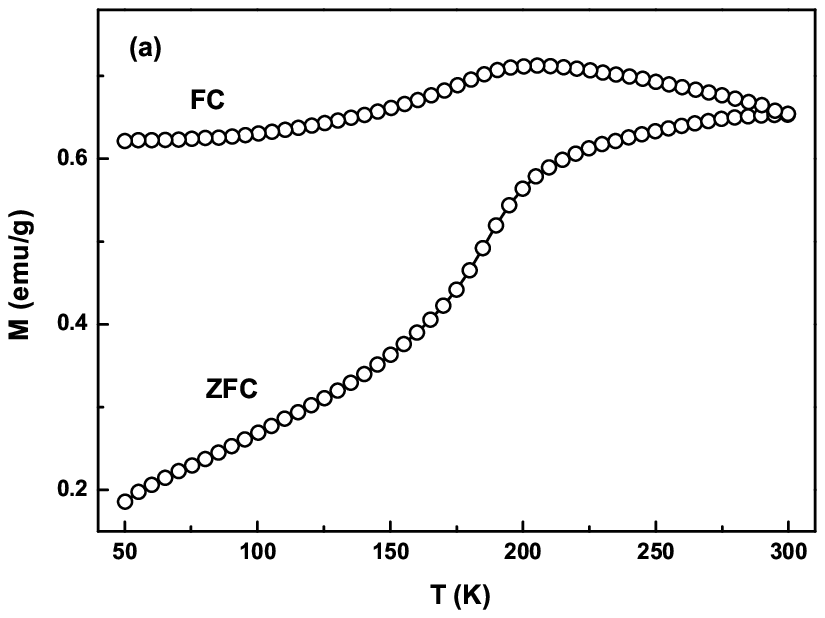} \\
    \includegraphics[width=7cm]{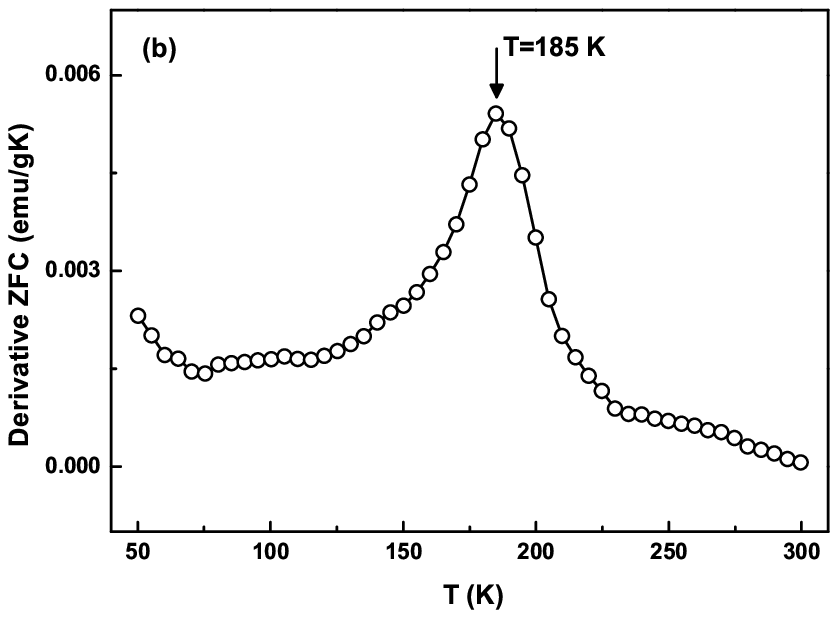}
    \end{center} 
\vspace{-.5cm}
    \caption{(a) Magnetization vs.\ temperature for both ZFC and FC magnetizations for the FeM40h sample and (b) derivative of the ZFC magnetization as a function of temperature.} 
    \label{Fig_04} 
\end{figure}

Figure~\ref{Fig_04} presents the magnetization behavior of the produced FeM40h sample as a function of the temperature. Figure~\ref{Fig_04}(a) shows the FC and ZFC magnetizations measured in the temperature range between $300$ K and $50$ K. Both ZFC and FC magnetization measurements exhibit peculiar features, mainly in the temperature interval between $175$ K and $200$ K. In this case, for the FC curve, a broad peak is observed in this region, evidenced by the increase of the magnetization signal as the temperature decreases from $300$ K to $200$ K, reaching the maximum value, and a subsequent decrease, until a nearly constant value below $60$ K. For the ZFC curve, the magnetization continuously decreases as the temperature goes from $300$ K to $50$ K, in which a more intense decrease of the $M$ value is verified in the range between $200$ K and $175$ K. 

The derivative of the ZFC curve, presented in Fig.~\ref{Fig_04}(b), indicates that the maximum of variation occurs at $185$ K, being this value close to the N\'{e}el temperature of the w\"{u}stite phase. For temperatures above $200$ K, the condition of derivative close to zero is satisfied indicating that the system becomes magnetically blocked at this temperature. The ZFC and FC curves separates at temperatures below $300$ K, it can be attributed to the existence of magnetic anisotropy barriers in the $\alpha$-Fe phase. It is noteworthy to mention that because w\"{u}stite is not stoichiometric, some ferro/ferrimagnetism is expected due to their uncompensated magnetic moments, thus the sudden increase at $200$ K in magnetic signal in the ZFC and FC curves could be related to this contribution. 

Small $\alpha$-Fe NPs use to exhibit anisotropy energies with one or two orders of magnitude larger~\cite{JAP103p07D521, JMMM331p156} than the bulk one, whose value is $K_{bulk} = 4.8 \times 10^4$ J/m$^3$. Thus, by using the well-known equation $T_B = KV/25K_B$, it is possible to estimate the blocking temperature $T_B$ from the magnetic anisotropy. In this case, considering $K \sim 5 \times 10^5$ J/m$^3$ and Fe NPs with diameters of $8$ nm, we obtain $T_B \sim 400$ K. In this sense, the calculated $T_B$ seems to indicate that the particles may be already blocked at $300$ K.

\begin{figure}[!h] 
    \begin{center} 
    \includegraphics[width=7cm]{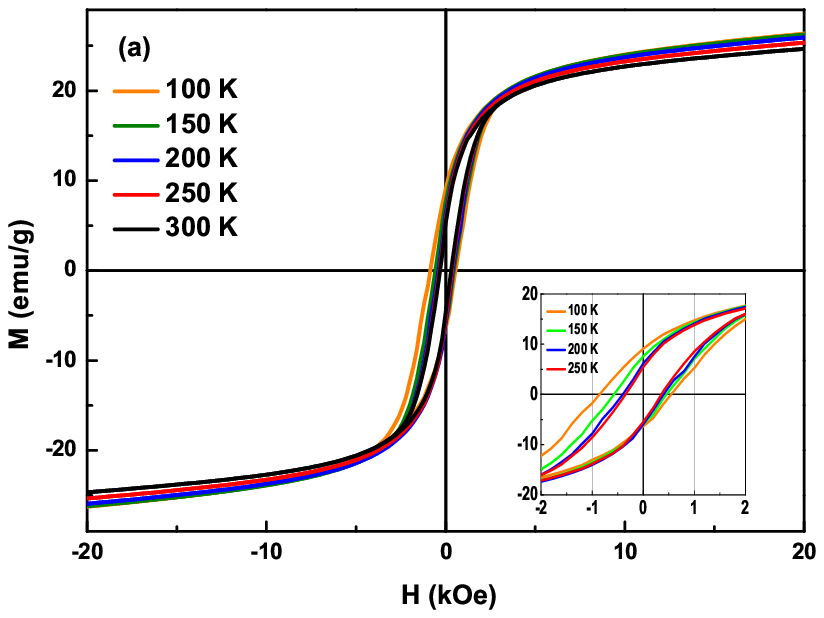} \\
    \includegraphics[width=7cm]{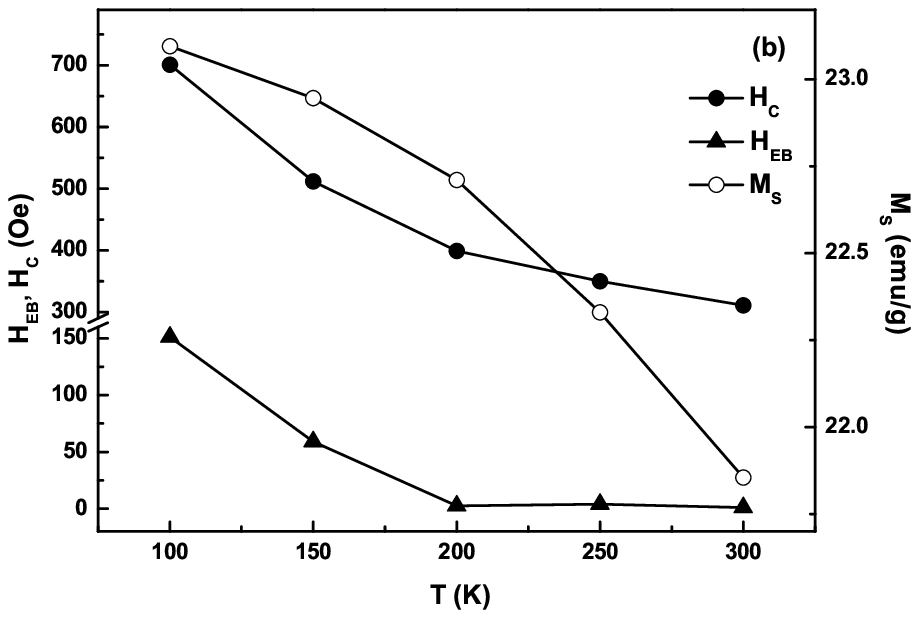}
    \end{center} 
\vspace{-.5cm}
    \caption{(a) DC field cooling magnetization curves acquired for the FeM40h sample for selected temperatures, in the interval between $300$ K and $100$ K. Although the plots are presented for a restricted magnetic field range, curves are obtained with maximum field of $30$ kOe. The inset shows in detail the shift of the curves due to exchange bias effect. (b) Exchange bias field, coercive field and saturation magnetization as a function of temperature.} 
    \label{Fig_05} 
\end{figure}

Figure~\ref{Fig_05} presents the magnetic properties of the produced FeM40h sample as a function of the magnetic field at selected temperatures. In this case, Fig.~\ref{Fig_05}(a) shows the DC field cooling magnetization curves acquired at the selected temperatures of $300$ K, $250$ K, $200$ K, $150$ K, and $100$ K. At low fields, it is possible to notice the appearance of the exchange bias effect as the temperature is reduced. All the hysteresis loops reach the saturation at magnetic field $\sim 5$ kOe. At higher fields, the small increase of magnetization is in agreement with the paramagnetic or antiferromagnetic behavior of the w\"{u}stite phase.
Figure~\ref{Fig_05}(b) presents the variations of the coercive field $H_C$ and exchange bias $H_{EB}$ magnetic field as a function of the temperature. In the case of coercivity, it is verified a noticeable increase as the temperature is reduced, from $H_C \sim 310$ to $\sim 700$ Oe. The exchange bias field quantifies the magnetic coupling effect between the phases and is defined by $H_{EB} = (H_{right} + H_{left})/2$, $H_{right}$ and $H_{left}$ being the points where the loop intersects the axis of the applied field. Since the exchange bias effect is observed and the magnetization curves are shifted, the coercive field is calculated according to $H_C = (H_{right} - H_{left})/2$. The field cooling hysteresis performed at $250$ K and $200$ K did not show exchange bias effect, but below $200$ K, the $H_{EB}$ values continuously increases. Thus, it is clear that the exchange bias effect is a result of the coupling of the ferromagnetic and antiferromagnetic materials composing the sample. 

At higher fields, $H > 20$ kOe, magnetization saturation $M_S$ can be calculated by fitting the magnetization data using the expression $M(H) = M_S + \chi H$, where $\chi$ is the magnetic susceptibility of the antiferromagnetic phase. From this point, considering the obtained $M_S$ value and assuming that the saturation magnetization for the $\alpha$-Fe is M$_{\alpha-\textrm{Fe}} =220$ emu/g~\cite{JAP39p669}, it is possible to calculate the $\alpha$-Fe mass contribution to the magnetic signal. Thus, Fig.~\ref{Fig_05}(b) also presents the changes of the saturation magnetization MS as a function of the temperature. In this case, an increase is also verified as the temperature decreases and $M_S$ reaches an equilibrium value of $\sim 23$ emu/g, at temperatures below $150$ K. In particular, for the temperature of $100$ K, the measured value $M_S = 23.09 \pm 0.02$ emu/g and $\chi = (1.65 \pm 0.01) \times 10^{-4}$ emu/gOe. The obtained $M_S$ value leads to a $\alpha$-Fe mass contribution to the magnetic signal of about $11$\%, value close to the ones found with the XRD and MS measurements. 

\subsection{Thermal characterization}
\label{Thermal_characterization}

Finally, in order to determine the magnetic power loss in the produced powder FeM40h sample, thermal characterization is performed via magnetic hyperthermia. It is known that the heating effect strongly depends on the magnetic particle's properties, measurement conditions, particle size distribution and particle dispersion, as well as on the experimental conditions~\cite{JMMM354p163}.

\begin{figure}[!h] 
    \begin{center} 
    \includegraphics[width=7cm]{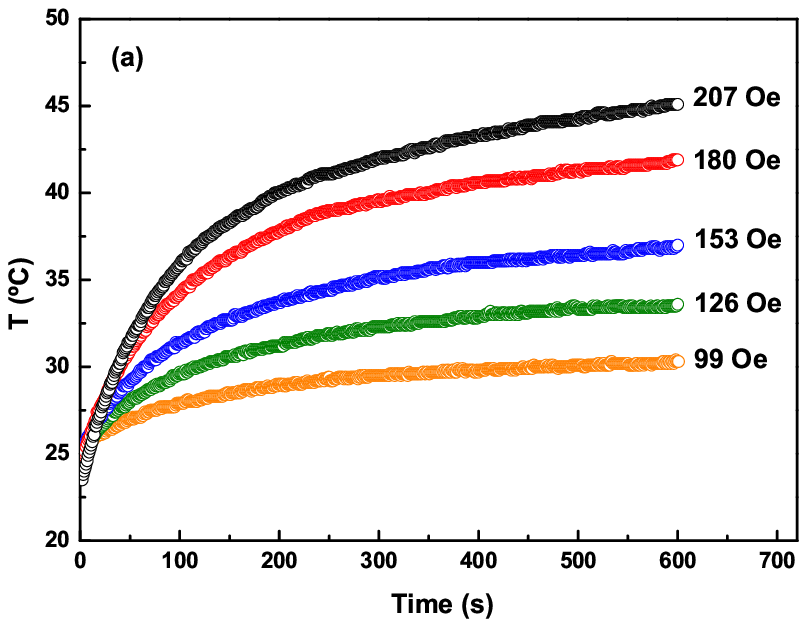} \\
    \includegraphics[width=7cm]{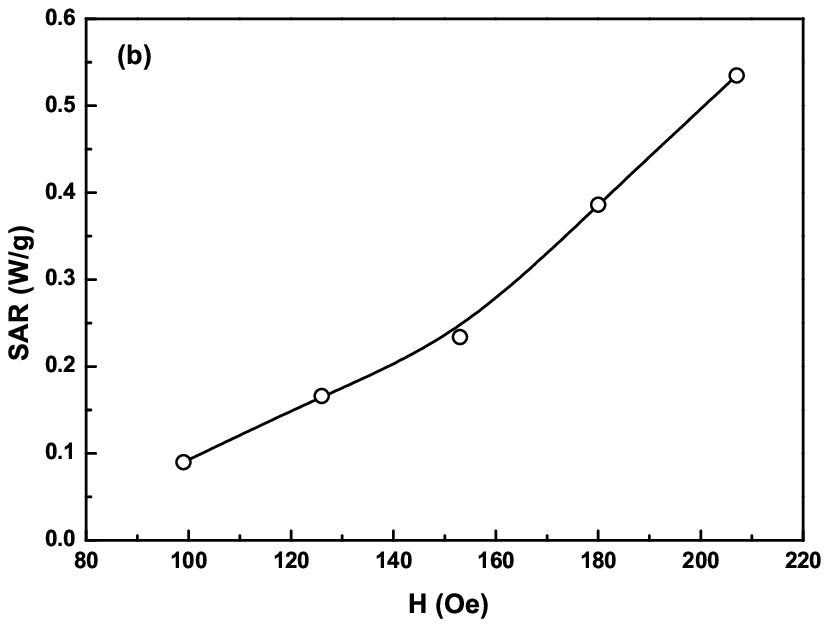}
    \end{center} 
\vspace{-.5cm}
    \caption{(a) Temperature of the produced FeM40h powder sample as a function of the time when submitted to an alternating magnetic field with frequency of $58$ kHz and selected field amplitudes, varying between $90$ to $210$ Oe. (b) Values of SAR as a function of the maximum amplitude of the alternating magnetic field, calculated from the temperature vs.\ time curves.} 
    \label{Fig_06} 
\end{figure}

Figure~\ref{Fig_06}(a) shows the plot of the temperature of the FeM40h sample as a function of the time in a magnetic hyperthermia experiment in which the external alternating magnetic field with frequency of $58$ kHz and maximum amplitude varying between $90$ to $210$ Oe is employed. In particular, it is verified a clear dependence of the reached temperature with the field amplitude, as expected. However, the most striking feature resides in the temperature reached when the sample is submitted to an alternating field with amplitude of $207$ Oe. In this case, the temperature increases up to $\sim 46\,^\circ$C in a short time interval, $\sim 600$ s. 

Thus, due to this reached temperature value, the time interval of temperature elevation, high stability of the Fe particles in time, the produced Fe nanoparticles embedded in a w\"{u}stite matrix powder becomes an interesting candidate for biological applications, such as the one employed for cancer therapy, since they are in principle suitable for heat ablation of cancer cells and the maximum value to preserve healthy tissues.

For purpose of discussion, here, we assume that the specific absortion rate (SAR), which is the amount of energy absorbed from an alternating magnetic field and converted to internal energy per time unit per Fe mass unit, verified through an increase of temperature of the sample, can be expressed as~\cite{JMMM268p33}:
\begin{equation}
\textrm{SAR} = c_{\textrm{Fe}} \frac{\Delta T}{\Delta t} \left (\frac{1}{m_{\textrm{Fe}}} \right ) \,,
\end{equation}
\noindent where $c_{\textrm{Fe}}$ is the iron's heat capacity, $\Delta T/\Delta t$ is the initial slope of the $T$ vs.\ curve $t$ presented in Fig.~\ref{Fig_06}(a), and $m_{\textrm{Fe}}$ is the iron content per gram of the whole sample. In this case, we consider the iron as the active material associated to the changes of temperature. 

Thus, Fig.~\ref{Fig_06}(b) presents the SAR values as a function of the maximum amplitude of the external alternating applied field. Typical values between $10$ and $100$ W/g are commonly found in literature for superparamagnetic nanoparticles immersed in an aqueous medium submitted to alternating magnetic fields up to $18$ kA/m. Besides, by using low particle concentration, higher field amplitude and higher frequency, very high SAR values can be reached. 

In our case, under the employed experimental conditions, the SAR values vary from $0.1$ to $0.5$ W/g. These values are in quantitative agreement with the ones obtained using similar experimental conditions for nanoparticle systems grown by evaporating targets of Fe and Fe$_3$O$_4$ with different stoichiometry~\cite{JAP114p103904}. The reduced values can be related to several factors. As a first point, high particle concentration is employed in the experiment, due to the considerable Fe mass. Second, since the field applied in the hyperthermia experiment is smaller than the coercive field of the sample, it is expected that the just part iron nanoparticles are responsible by the magnetization changes in this field interval, i. e., a considerable portion of the Fe NPs are not contributing to the sample heating. Third, N\'{e}el switching mechanism can not be considered because the nanoparticles are not in the superparmagnetic regime, as well as the Brownian mechanism is largely irrelevant, since our sample is a powder non-immersed in an aqueous medium~\cite{JMMM354p163}. Anyway, despite the SAR values, very interesting temperatures can be reached in a reduced time interval, similarly to several results find in literature~\cite{JMMM354p163, JN2013p181820}. 

\section{Conclusion}
\label{Conclusion}

In summary, in this paper we performed a systematic study of the structural, magnetic and thermal properties of a powder sample of Fe nanoparticles passivated by a w\"{u}stite matrix. In particular, we are able to prepare nanoparticles with interesting properties by using high energy mechanical milling from iron powder and water. The produced nanoparticles present high stability in time. Besides, exchange bias interaction is observed below the N\'{e}el temperature of the w\"{u}stite. Finally, when taken into account the thermal properties, magnetic hyperthermia experiments present promising results, in which the Fe nanoparticles embedded in a w\"{u}stite matrix reaches temperatures of $\sim 46\, ^\circ$C in a short time interval, becoming an interesting candidate for biological applications, such as the one employed for cancer therapy. These results correspond to a step to understand the complex behavior of nanoparticles submitted to an alternating magnetic field. Considering this fact, in order to obtain a complete general framework on this issue, more experimental and theoretical studies are needed, including magnetic hyperthermia experiments considering higher magnetic field and distinct field frequencies. These experiments and analyses are currently in progress.

\begin{acknowledgments} 
The research is supported by the Brazilian agencies CNPq (Grants No.~$471302$/$2013$-$9$, No.~$310761$/$2011$-$5$, No.~$555620$/$2010$-$7$, and No.~$305234$/$2012$-$9$), CAPES, and FAPERN (Grant Pronem No.~$03/2012$). F.B. acknowledges financial support of the INCT of Space Studies. 
\end{acknowledgments}

\end{document}